# Astrocytes: orchestrating synaptic plasticity?

Maurizio De Pittà[1,2,*], Nicolas Brunel[1,3], Andrea Volterra[4,*]

1. Department of Neurobiology, The University of Chicago, Chicago, IL, USA.

2. EPI BEAGLE, INRIA Rhône-Alpes, Villeurbanne, France.

3. Department of Statistics, The University of Chicago, Chicago, IL, USA.

4. Département de Neurosciences Fondamentales, Université de Lausanne, Switzerland.

∗ Corresponding Authors:

Maurizio De Pittà:

Department of Neurobiology, The University of Chicago

947 East 58th Street MC 0928, Room AB320; Chicago, IL 60637, USA

Tel.: +1 773 702 6606

Fax: +1 773 702 3774

mauriziodepitta@gmail.com

Andrea Volterra :

Département de Neurosciences Fondamentales, Université de Lausanne

Rue du Bugnon, 9 ; 1005 Lausanne, Switzerland

Tel. : +41 21 692 5271

Fax : +41 21 692 5105 (or 692 5255)

Andrea.Volterra@unil.ch

**Abstract**

Synaptic plasticity is the capacity of a preexisting connection between two neurons to change in strength as a function of neural activity. Because synaptic plasticity is the major candidate mechanism for learning and memory, the elucidation of its constituting mechanisms is of crucial importance in many aspects of normal and pathological brain function. In particular, a prominent aspect that remains debated is how the plasticity mechanisms, that encompass a broad spectrum of temporal and spatial scales, come to play together in a concerted fashion. Here we review and discuss evidence that pinpoints to a possible non-neuronal, glial candidate for such orchestration: the regulation of synaptic plasticity by astrocytes.

## Keywords

**Abbreviations**

A1R: adenosine A1 receptor; AMPAR: α-amino-3-hydroxy-5-methyl-4-isoxazolepropionic acid receptor; CB1R: cannabinoid receptor type 1; EPSC: excitatory postsynaptic current; GPCR: G protein coupled receptor; LTD: long-term depression; LTP: long-term potentiation; NMDAR: N-methyl-D-aspartate receptor; STDP: spike-timing—dependent plasticity; tLTD: spike-timing—dependent long-term depression; tLTP: spike-timing—dependent long-term potentiation; TNFα: tumor necrosis factor alpha.

**Introduction**

A fundamental property of synapses is that they transmit signals between neurons in an ever-changing manner. In particular, the effect of a synaptically transmitted signal from one neuron to another neuron can vary greatly depending on the history of activity at either or both sides of the synapse, and such variations may last from milliseconds to months (Citri and Malenka, 2008). Activity-dependent changes of synaptic transmission arise from a large number of mechanisms, collectively known as synaptic plasticity. Depending on the time scale of these mechanisms, synaptic plasticity can be divided into three broad categories: (1) short-term plasticity, which refers to changes that occur over milliseconds to minutes, like for example depression or facilitation of neurotransmitter release, allowing synapses to perform critical computational functions in neural circuits (Abbott and Regehr, 2004); (2) long-term plasticity, which involves changes that may last for hours up to days, weeks or even months (Abraham, 2003), and includes the classic Hebbian plasticity in the form of long-term potentiation (LTP) and depression (LTD) that are thought to underpin learning and memory (Martin et al., 2000); (3) homeostatic plasticity, which may regard both synapses and neurons and allows neural circuits to maintain appropriate levels of excitability and connectivity despite changes in the surrounding environment brought about by metabolism and experience-dependent plasticity (Turrigiano, 2011).

Synaptic plasticity is typically input-specific (or homosynaptic), meaning that stimulation of particular input(s) to a neuron alters the efficiency of the activated synaptic connection(s) (Bear and Malenka, 1994). However, plasticity may also occur heterosynaptically whenever the stimulation of a particular neuron may lead to changes in the strength of synaptic connections between input pathways that have not been stimulated and the stimulated neuron (Bailey et al., 2000). Both short- and long-term forms of synaptic plasticity may occur homosynaptically and heterosynaptically (Malenka and Bear, 2004), whereas homeostatic plasticity has mostly been documented as a global mechanism, occurring across the entire population of a neuron's synapses (Turrigiano, 2008), although recent evidence hints it could act on individual or small groups of synapses too (Vitureira et al., 2012).

A fundamental issue in the study of synaptic plasticity is how neural circuits undergoing plastic changes maintain stability and function (Watt and Desai, 2010). Associative (Hebbian) plasticity resulting from correlated pre- and postsynaptic firing activity generates a positive feedback process (Dayan and Abbott, 2001). The activity that strengthens (weakens) synapses is reinforced (lessened) by Hebbian plasticity, which leads to more (less) activity and to further synaptic modification. In this fashion, unless changes in synaptic strength across multiple synapses are coordinated appropriately, the level of activity in a neural circuit could grow or shrink in an uncontrolled manner, while individual neurons could lose selectivity to different patterns of input (Abbott and Nelson, 2000). This scenario could be avoided by making synapses compete for control of postsynaptic firing activity, while keeping the average synaptic strength under control. Different mechanisms have been proposed to achieve such a synaptic competition in theoretical models of synaptic plasticity. In the Bienenstock-Cooper-Munro (BCM) rule (Bienenstock et al., 1982), this competition is introduced thanks to a metaplasticity mechanism, whereby the threshold postsynaptic rate that separates LTP from LTD induction moves based on prior activity (Abraham, 2008; Abraham and Bear, 1996). Hence, if the average firing rate of a postsynaptic neuron increases, its threshold for LTP will increase accordingly, potentially leading to a stabilization of the average firing rate, while maintaining selectivity. Alternatively, in spike-timing—dependent plasticity (STDP) rules (Abbott and Nelson, 2000), synapses compete to drive postsynaptic firing. The ones that manage to drive postsynaptic firing get strengthened, while the others get weakened, provided that depression is on average stronger than potentiation. This ultimately leads to a nonuniform distribution of synaptic weights that drives neurons to a noisy but temporally sensitive firing state, consistent with in vivo observations (Shadlen and Newsome, 1994).

Although in principle metaplasticity might regulate the total synaptic drive if the effects of LTP can be counteracted by those of LTD (and vice versa), in practice this requires a rather delicate balance between the two that is difficult to achieve for rate-based plasticity (Gerstner and Kistler, 2002; Toyoizumi et al., 2014a; Zenke et al., 2014). STDP could allow for such balance but only if specific constraints such as the dominance of spike-timing—dependent long-term depression (tLTD) over potentiation (tLTP) are fulfilled (Song et al., 2000). Overall, while balancing LTP and LTD should be useful for homeostasis, present evidence suggests that it is best complemented by additional mechanisms of plasticity, possibly acting on multiple spatial and temporal scales (Desai, 2003; Watt and Desai, 2010). Consequently, a point of particular importance is to understand how different spatial and temporal scales of plasticity could come to play in a concerted fashion, and how their coexistence could benefit neural network dynamics, and its capacity to process, learn and store information (Nelson and Turrigiano, 2008).

Theoretical studies have provided insightful predictions on the functional relevance of some forms of synaptic plasticity like synaptic scaling and redistribution (Gray et al., 2006; Pozo and Goda, 2010a). For example, homeostatic scaling of all synapses of a neuron, which is obtained by adjusting the number of postsynaptic receptors proportionally to the neuron's average firing activity (Lissin et al., 1998; O'Brien et al., 1998; Turrigiano et al., 1998), could maintain neural activity within a dynamic range and, more generally, stabilize neural circuit function despite the positive feedback of Hebbian plasticity (Toyoizumi et al., 2014b). Similarly, the presynaptic redistribution of synaptic strength, obtained at cortical synapses via an increase (decrease) of the probability of transmitter release in concomitance with LTP (LTD) (Markram and Tsodyks, 1996a; Sjöström et al., 2003; Volgushe et al., 1997; Yasui et al., 2005), would allow Hebbian plasticity to act without increasing postsynaptic firing rates or network excitability at steady state (Abbott and Nelson, 2000).

Independently of the spatiotemporal scale under consideration, synaptic plasticity has been generally regarded as an inherent property of synapses, that is a property based on mechanistic changes that develop and occur within neurons (Cooke and Bliss, 2006). However, in recent years, a growing body of evidence has suggested that proper synaptic functioning may involve an active participation of astrocytes, the main type of glial cells (Auld and Robitaille, 2003; Pannasch and Rouach, 2013; Volterra and Meldolesi, 2005), thus implying that synaptic plasticity itself could, to some extent, be under astrocytic regulation (Allen and Barres, 2005; López-Hidalgo and Schummers, 2014; Perea et al., 2009a). The present review summarizes the evidence for an astrocyte involvement in synaptic plasticity, discussing its potential role(s) in neural network function and stability.

## Synapse-astrocyte coupling

Beyond a recognized role in the genesis and elimination of synapses (reviewed by Eroglu and Barres (2010)), astrocytes in the mature brain often surround neuronal somata and dendrites, and provide fine ensheathment of synapses (Chao et al., 2002). Although the extent of the astrocytic ensheathment largely varies with the brain region, hinting at local specializations (Theodosis et al., 2008a), in the rodent brain a single astrocyte could cover hundreds of thousands of synapses (Bushong et al., 2002; Halassa et al., 2007), and these figures could be one order of magnitude larger in humans (Oberheim et al., 2009). Such morphological arrangement provides the structural substrate for tight functional interactions between astrocytes and neurons (Bernardinelli et al., 2014a; Saab et al., 2012; Theodosis et al., 2008b).

Thanks to their strategic position, astrocytes are indeed known to contribute to the regulation of the neuronal microenvironment by maintaining a tight control on local ion (Simard and Nedergaard, 2004) and pH homeostasis (Deitmer, 2004), delivering metabolic substrates to neurons (Brown and Ransom, 2007), as well as controlling the microvasculature (Attwell et al., 2010) and clearing away metabolic waste (Nedergaard, 2013). Furthermore, the perisynaptic processes of astrocytes can act as physical barriers for spillover and diffusion into the extrasynaptic space of locally released, potentially active molecules, limiting crosstalk between neighboring synapses while favoring specificity of synaptic transmission (Beenhakker and Huguenard, 2010; Oliet et al., 2001; Piet et al., 2004; Tanaka et al., 2013; Ventura and Harris, 1999). Perisynaptic astrocytic processes are indeed enriched in transporters that assure rapid and efficient removal of synaptically-released neurotransmitters, in particular glutamate (Anderson and Swanson, 2000) and GABA (Conti et al., 2004). The control of the speed and the extent of neurotransmitter clearance by astrocytes could also have a role in synaptic plasticity inasmuch as it affects the degree of postsynaptic activation and desensitization (Bergles and Rothstein, 2004; Tzingounis and Wadiche, 2007).

Besides transporters however, astrocytes express a variety of receptors for typical neurotransmitter molecules like glutamate, acetylcholine, ATP, GABA, norepinephrine, and for retrograde messengers such as endocannabinoids (Charles et al., 2003; Haydon, 2001; Porter and McCarthy, 1997), whereby they sense synaptic activity and display, in response, transient elevations of their intracellular (cytosolic) $Ca^{2+}$ concentration (Zorec et al., 2012). Calcium activation of astrocytes is expected to continuously occur in the living brain, as it has been observed in vivo in different brain areas, in conditions of physiological, local synaptic activity (Volterra et al., 2014) as well as in response to mechanic (Lind et al., 2013; Perez-Alvarez et al., 2014; Takata et al., 2011; Wang et al., 2006; Winship et al., 2007) and visual sensory stimulation (Chen et al., 2012; Schummers et al., 2008) and motor activation (Dombeck et al., 2007; Nimmerjahn et al., 2009; Paukert et al., 2014). Moreover it could also occur in response to endogenous activity (Hirase et al., 2004; Hoogland et al., 2009; Kuga et al., 2011; Takata and Hirase, 2008).

What is the functional significance of such $Ca^{2+}$ signaling? What downstream responses does it trigger? A potential class of responses that have elicited great interest in recent years, is the release of gliotransmitters from the astrocyte (Newman, 2003). Gliotransmitters are neuroactive molecules such as glutamate, ATP, GABA, D-serine or the cytokine tumor necrosis factor alpha (TNFα), that are so-called for their glial origin as opposed to neurotransmitters (Bezzi and Volterra, 2001). These gliotransmitters are released by astrocytes into the extracellular space, where they may diffuse to neuronal elements and promote further downstream processes, a phenomenon termed gliotransmission (Araque et al., 2014). Two pathways are particularly relevant for their possible implications on synaptic functions: the gliotransmitter-mediated activation of extrasynaptic receptors located respectively on presynaptic and postsynaptic terminals (Araque et al., 2014; Halassa and Haydon, 2010; Perea et al., 2009b; Volterra and Meldolesi, 2005) (Figure1). Remarkably, these two pathways could provide activity-dependent feedback and feedforward interactions between pre- and postsynaptic terminals, making the flow of input action potentials to postsynaptic responses through synapses no more just unidirectional, due to the additional astrocytic component (De Pittà et al., 2012).

How effectively do these gliotransmitter-mediated pathways come into play in physiological synaptic transmission? Neither the molecular mechanisms of gliotransmitter release (critically reviewed by Sahlender et al. (2014)), nor their regulation is yet fully understood. With some possible exceptions, like in the cerebellum (Brockhaus and Deitmer, 2002), $Ca^{2+}$ is believed to be the main regulator of gliotransmission and the fact that astrocytic $Ca^{2+}$ signals occur on a physiological basis hints that it is also true for gliotransmission (Navarrete and Araque, 2011). A major question is how astrocytic $Ca^{2+}$ responses, which have been reported to be much slower than their neuronal homologous (ranging from hundreds of millisecond to tens of seconds) interact with the (much faster) synaptic dynamics (Agulhon et al., 2012; Haydon, 2001). To understand better this interaction, one first needs to properly identify the input and output signals of the astrocyte with respect to its possible effects on synaptic transmission. From this perspective, both $Ca^{2+}$ and the gliotransmitter signals could be regarded just as mediators of the astrocyte effect on the synapse, while the effective "readout" for the astrocyte effect could rather be identified in the activation of extrasynaptic neuronal receptors targeted by gliotransmitters. This allows focusing the discussion that follows on the time course of the downstream signaling triggered by these activated receptors, which is generally measurable in experiments (Araque et al., 1998a; Perea and Araque, 2007; Perea et al., 2014), without the need to consider in detail the kinetics of gliotransmitter release. It shall be kept in mind nonetheless, that specific conditions might be required for gliotransmission to activate extrasynaptic receptors (Pascual et al., 2011; Santello et al., 2011), and this activation is generally not temporally coincident with synaptic transmission but rather exerts a modulatory control on the latter (Araque et al., 2014).

**Astrocyte regulation of short-term synaptic plasticity**

Activation of presynaptic receptors by astrocytic gliotransmitters may trigger different receptor-specific downstream signaling pathways (Engelman and MacDermott, 2004; Haas and Selbach, 2000; Pinheiro and Mulle, 2008), yet all pathways ultimately seem to share a common readout that is the modulation of the probability of synaptic release (Miller, 1998). The short-lived version of this gliotransmitter-mediated modulation lasts for tens of seconds (Fiacco and McCarthy, 2004; Jourdain et al., 2007) to few minutes (Perea and Araque, 2007; Perea et al., 2014; Perez-Alvarez et al., 2014). Hence it acts on much longer time scales than those of typical processes involved in synaptic release like vesicle release and reintegration, which are on time scales of hundreds of microseconds and milliseconds respectively (Schneggenburger and Neher, 2000), being a potential candidate in the regulation of form of short-term synaptic plasticity like short-term depression and facilitation. In this respect, this astrocyte-mediated regulation resembles neuromodulation of synaptic release (Hirase et al., 2014), with the important difference that, while the latter was proposed to essentially pertain to neuronal signaling by volume transmission, gliotransmission could also occur in a focal fashion (Santello and Volterra, 2009), particularly when involving release of glutamate, whose extracellular levels are tightly regulated by uptake (Jourdain et al., 2007). This in turn, could produce spatially confined and temporally precise modulations of synaptic transmission (De Pittà et al., 2012).

Depending on receptor type, the modulation of synaptic release probability by gliotransmitter-activated presynaptic receptors may be either toward an increase or toward a decrease of the frequency of spontaneous (Bonansco et al., 2011; Di Castro et al., 2011; Fiacco and McCarthy, 2004; Jourdain et al., 2007; Panatier et al., 2011; Pascual, 2005; Perea et al., 2014) and evoked neurotransmitter release, both in excitatory (Halassa et al., 2009; Jourdain et al., 2007; Navarrete and Araque, 2010; Panatier et al., 2011; Perea and Araque, 2007; Perez-Alvarez et al., 2014; Schmitt et al., 2012; Zhang et al., 2003) and inhibitory synapses (Benedetti et al., 2011; Kang et al., 1998; Liu et al., 2004a, 2004b Brockhaus and Deitmer, 2002). As a result, synapses whose release probability is increased by astrocytic gliotransmitters, show a reduction in their paired-pulse ratio (Bonansco et al., 2011; Jourdain et al., 2007), that is the ratio of the second over the first postsynaptic current triggered by a pair of properly-timed pulses delivered presynaptically (Zucker and Regehr, 2002). On the contrary, synapses whose release probability is decreased by gliotransmission are prone to display an increase of the paired-pulse ratio (Liu et al., 2004b). These results can be explained recalling that synaptic release probability dictates how fast synaptic neurotransmitter resources are to be depleted (Zucker and Regehr, 2002). Thus, synapses with release probability increased by gliotransmission are likely to run out faster of neurotransmitter, exhibiting rapid onset of short-term depression and thus lower values of paired-pulse ratio. On the contrary, synapses whose release probability is reduced by gliotransmission, deplete their neurotransmitter resources slower and may exhibit frequency-dependent facilitation consistent with large values of paired-pulse ratio (Dittman et al., 2000).

Both glutamate and purine presynaptic receptors are known targets of gliotransmitters, yet their differential recruitment likely depends on developmental, regional and physiological factors (reviewed in De Pittà et al. (2012)). Several experiments by a few groups have however questioned the capacity of astrocytes to induce gliotransmitter-dependent synaptic modulation on a physiological basis (Agulhon et al., 2010; Fiacco et al., 2007; Lovatt et al., 2012; Petravicz et al., 2008). Indeed both stimulation (Agulhon et al., 2010; Fiacco et al., 2007) and removal by genetic tools (Petravicz et al., 2008) of inositol 1,4,5-trisphospate (IP$_3$)-dependent astrocytic Ca$^{2+}$ signaling, thought to be the main trigger of gliotransmitter release, were reported by these groups to not affect synaptic transmission. In parallel, the astrocytic origin of purinergic synaptic modulation has been questioned in light of a possible neuronal origin for the purines responsible for it (Fujita et al., 2014; Lovatt et al., 2012). While several explanations for these seemingly contradictory results exist (Araque et al., 2014; Nedergaard and Verkhratsky, 2012; Volterra et al., 2014), overall they suggest that synaptic modulation by gliotransmitters is not a straightforward process, but rather likely requires specific conditions to occur (Araque et al., 2014). In particular, one may predict that, to detect a change in synaptic release probability mediated by the astrocyte, a sufficient number of presynaptic receptors must be recruited for a sufficient time by gliotransmitter molecules, which would only be possible if enough gliotransmitter is made available extracellularly by the astrocyte. Theoretical calculations support the possibility that a single quantal release of gliotransmitter, like glutamate for example, is sufficient to activate presynaptic receptors (Hamilton and Attwell, 2010). Yet, possibly many quanta of glutamate are released at once in a burst-like fashion to allow this transmitter escape reuptake by astrocytic transporters and functionally activate presynaptic receptors (Santello et al., 2011). On the other hand, once released into the extracellular space, glutamate, and so other gliotransmitters, may also be cleared by diffusion (Montana et al., 2006) and/or enzymatic degradation (Abbracchio et al., 2009). To counteract this possibility, and guarantee activation of presynaptic receptors, gliotransmitter release from the astrocyte – independently of the underlying mechanism (single or multi-quantal or other) – must occur at sufficiently high rate. Recent experiments (Perez-Alvarez et al., 2014) indeed support this scenario, hinting that the entity of modulation of synaptic release probability scales as the number of astrocytic $Ca^{2+}$ spikes, which likely correlates with the number of gliotransmitter release events (Marchaland et al., 2008; Pasti et al., 2001a). In this respect, it must be emphasized that is not just the entity but also the frequency of astrocytic $Ca^{2+}$ elevations that can dictate the number of gliotransmitter release events and therefore their synaptic efficacy (Araque et al., 2014; Volterra et al., 2014).

Theoretical arguments support the existence of a threshold in the rate of gliotransmitter release able to induce a change of synaptic plasticity and allow the identification of a number of conditions necessary to reach such a threshold (Modelling Box). These conditions are illustrated in Figure 2 where the threshold rate is color mapped as a function of the resting release probability ($b$) of the synapse and the ratio between the time constants for facilitation and depression ($\tau_f / \tau_d$). In these maps, a change of short-term plasticity from predominant frequency-dependent facilitation to depression (or vice versa) corresponds to an increase (decrease) of $b$ beyond the white-dashed threshold at fixed $\tau_f / \tau_d$ (Figures 2A,B). It may be seen in particular that three factors could critically control whether or not a synapse could be effectively modulated by astrocytic gliotransmitters: (1) the capacity of the astrocyte for efficient gliotransmission, as reflected by a rate of reintegration of releasable gliotransmitter resources sufficiently fast to avoid their use-dependent depletion (Figure 2C); (2) the type of the presynaptic receptors available and, in particular, the dynamics of their downstream signaling responsible for synaptic release modulation: the slower this signaling evolves with respect to the synapse's time scale of short-term plasticity, the more detectable the modulation is (Figure 2D); and (3) how likely the gliotransmitter is to reach and activate presynaptic receptors, considering the receptors' affinity to bind gliotransmitters, clearance of the gliotransmitters in the extracellular space, and how easily accessible presynaptic receptors are with respect to the site of gliotransmitter release (Figure 2E). Thus, if a perisynaptic astrocytic process is not sufficiently close to the presynaptic terminal, one may expect that the majority of gliotransmitter is cleared away before it reaches presynaptic receptors. This could explain why, for example, gliotransmitter-mediated short-term potentiation of hippocampal synapses is no longer observed after LTP-induced retraction of perisynaptic astrocytic processes (Perez-Alvarez et al., 2014). On the other hand, in some hippocampal or cerebellar regions, where astrocytic ensheathment of synapses was reported to be mainly postsynaptic (Lehre and Rusakov, 2002), dynamic reshaping of astrocytic processes could be speculated to allow also for the opposite phenomenon: i.e., the appearance and fine tuning of gliotransmitter-mediated modulations of short-term plasticity in an activity-dependent fashion (Bernardinelli et al., 2014b; Lavialle et al., 2011).

  The interplay between multiple forms of short-term plasticity has profound influence on synaptic strength and critically shapes how synapses filter and transmit action potentials (Dittman et al., 2000; Fortune and Rose, 2001; Markram et al., 1998a). Remarkably, the filtering characteristics of a given synapse may not be fixed but rather be adjusted through modulation of the basal release probability by astrocytic gliotransmitters. An interesting possibility is that gliotransmitters tonically activate presynaptic receptors, thus setting the probability for basal synaptic release (Navarrete and Araque, 2011) and thereby, the response properties of the synapse (Markram et al., 1998a). This in turn suggests that the very nature of a synapse assessed by conventional electrophysiological experiments, i.e. whether it is mostly depressing or facilitating, could ultimately depend on the ongoing activation of presynaptic receptors by astrocytic gliotransmitters. Figure 3 exemplifies this principle in a standard synapse model (Mongillo et al., 2008; Tsodyks, 2005) adapted to include astrocytic gliotransmission (De Pittà et al., 2011). Hippocampal synapses, like for example Schaffer collateral synapses (Dittman et al., 2000), are often characterized by intermediate values of release probability, and act as band-pass filters due to the interplay of frequency-dependent facilitation and depression (Figure 3A). Notably, these synapses are most effective in transmitting action potentials for intermediate rates of presynaptic activity (Figure 3C, *black* PSC trace). However, at low-to-intermediate rates, when facilitation is prominent, the time-average postsynaptic response reflects an attenuated version of the integral of the stimulus, while at high rates of incoming action potentials, the response is mostly sensitive to rate variations (i.e. the derivative of the stimulus) (Tsodyks, 2005). This behavior could dramatically change, however, in the presence of synaptic potentiation by gliotransmitters as observed experimentally (Jourdain et al., 2007; Navarrete and Araque, 2010; Panatier et al., 2011). In this case, for a sufficiently large increase of release probability mediated by gliotransmitters, the synapse could turn into depressing and, act essentially as low-pass filter (Figure 3B*)* (Galarreta and Hestrin, 1998; Varela et al., 1997). Thus, stimuli that were originally integrated or simply buffered become instead respectively buffered or derived (Figure 3C, *red* PSC trace). This suggests that a hippocampal synapse could process the same stimulus in different ways depending on the regulation by astrocytic gliotransmission. This additional regulation mechanism could potentially increase the computational power of neural circuits in which such regulation operates.

### Astrocyte-regulated long-term synaptic plasticity

Several experiments in situ have shown that gliotransmission-mediated activation of presynaptic metabotropic glutamate receptors at hippocampal synapses and presynaptic NMDA receptors (NMDARs) at cortical synapses leads respectively to an increase (Navarrete et al., 2012; Perea and Araque, 2007) or decrease of synaptic release probability (Min and Nevian, 2012) that lasts tens of minutes, hinting at an astrocyte-mediated presynaptic mechanism for long-term plasticity. Such long-lasting modulations of synaptic release require postsynaptic stimulation in concomitance with presynaptic and astrocytic $Ca^{2+}$ stimulation (Perea and Araque, 2007), and seem to be mediated by retrograde endocannabinoid signaling (reviewed in Navarrete et al. (2014)). Intriguingly, long-term NMDA-dependent depression of synaptic release mediated by astrocytes is apparently necessary and sufficient for endocannabinoid-mediated spike-timing—dependent depression (tLTD) at barrel cortex synapses between excitatory neurons of layer 4 and layer 2/3 (Min and Nevian, 2012). tLTD at these synapses is known to require cooperative activation of CB1 receptors (CB1Rs) and presynaptic NMDARs (Bender et al., 2006; Corlew et al., 2007) but the mechanistic basis of this requirement is not known (Heifets and Castillo, 2009; Rodríguez-Moreno and Paulsen, 2010). Involvement of astrocytes could explain it inasmuch as, by expressing CB1Rs, they respond to postsynaptically-derived endocannabinoids by intracellular $Ca^{2+}$ elevation, release in turn glutamate, and mediate thereby activation of presynaptic NMDARs necessary for tLTD induction (Min and Nevian, 2012). This scenario opens to the possibility that astrocytes are critically involved in barrel cortex sensory map plasticity, for which tLTD is regarded as a key mechanism during development (Feldman, 2009; Li et al., 2009).

The other pathway whereby gliotransmitters could mediate long-term plasticity is postsynaptic, via regulation of either the number or the efficacy of receptor channels (Bains and Oliet, 2007). In behaving mice, for example, the decrease of AMPA receptors (AMPARs) by cannabinoid-induced LTD at hippocampal glutamatergic synapses, requires CB1R expression in astrocytes and CB1R-mediated glutamatergic gliotransmission (Han et al., 2012). In the hypothalamic paraventricular nucleus instead, adrenergic stimulation of ATP release from astrocytes results in LTP of glutamatergic synapses onto magnocellular secretory neurons (Gordon et al., 2005). This pathway is mediated by postsynaptic P2X channels which likely promote AMPAR insertion through activation of phosphatidylinositol 3-kinase (Gordon et al., 2005), akin of LTP occurring in other brain regions (Baxter and Wyllie, 2006; Kelly and Lynch, 2000; Qin et al., 2005; Raymond et al., 2002). Finally, in the immature cerebellum, activity-dependent LTD of synapses between parallel fiber and Purkinje neurons requires D-serine release from Bergmann glia, astrocyte-like cells of this region (Kakegawa et al., 2011). In this specific system, D-serine activates postsynaptic δ2 glutamate receptors, thereby causing internalization of AMPARs and thus LTD.

The involvement of D-serinergic gliotransmission in the mediation of long-term plasticity is not restricted to the cerebellum, however, although D-serine may act via different mechanisms in different areas or at different developmental stages (Bains and Oliet, 2007). Notably, D-serine, most likely secreted from astrocytes, is recognized as one of the major co-agonists of postsynaptic NMDARs at many excitatory synapses (Mothet et al., 2000; Papouin et al., 2012; Schell et al., 1995; Yang et al., 2003). In this fashion, by controlling the level of activation of these receptors, D-serinergic gliotransmission could critically mediate NMDAR-dependent LTP (Mothet et al., 2006; Panatier et al., 2006; Yang et al., 2003) and LTD (Zhang et al., 2008). D-serine release from astrocytes was indeed shown to be necessary for LTP of synapses in the hippocampus (Henneberger et al., 2010; López-Hidalgo et al., 2012; Yang et al., 2005, 2003), hypothalamus (Panatier et al., 2006), prefrontal (Fossat et al., 2012) and sensory cortex (Takata et al., 2011).

The hitherto reviewed studies suggest that at synapses, most notably in the hippocampus, multiple gliotransmitter pathways (e.g. glutamate and D-serine) could coexist, potentially underpinning different plasticity mechanisms (Henneberger et al., 2010; López-Hidalgo et al., 2012; Navarrete et al., 2012; Perea and Araque, 2007). What are then the possible conditions for the occurrence of one mechanism with respect to the other? Although several arguments suggest a dependence on functional and morphological specificity of the coupling of synaptic terminals with perisynaptic astrocytic processes (Araque et al., 2014), the very nature of the plasticity-inducing stimulus could also be crucial (Ben Achour et al., 2010; Min et al., 2012). Different patterns of synaptic activity could indeed release different factors that activate astrocytes in a different manner (Perea and Araque, 2005), thus resulting in different $Ca^{2+}$ patterns that could associate with different modes of gliotransmission (Pasti et al., 2001b; Shigetomi et al., 2008). Recent experiments in vivo showed, for example, that induction of LTP by cholinergic innervations in the hippocampus (Navarrete et al., 2012), somatosensory (Takata et al., 2011) and visual cortex (Chen et al., 2012), is subordinated to the stimulation of astrocytic $Ca^{2+}$ signaling by these very innervations. However, because cholinergic afferents to each of these brain regions are different, this could possibly account for the different gliotransmitters mediating such LTP: D-serine in the somatosensory cortex (Takata et al., 2011) and glutamate in the hippocampus (Navarrete et al., 2012).

The essential requirement for Hebbian plasticity is associativity, that is the correlation between pre- and postsynaptic activities (Gerstner and Kistler, 2002; Hebb, 1949). Remarkably, participation of gliotransmission to the onset of plasticity, as seen in all the above studies, makes synaptic associativity a condition necessary but no longer sufficient, since astrocytic $Ca^{2+}$ activation would also be needed for long-term plastic changes of synaptic strength. Intriguingly, the inclusion of astrocyte signaling in Hebbian plasticity is expected to add nonlinearity to associativity. Associativity depends on postsynaptic membrane potential (Lisman and Spruston, 2005; Sjöström et al., 2008) whose value is a function – among other factors – of the timing and amplitude of EPSPs (Caporale and Dan, 2008), and these latter could be modulated by gliotransmission, acting either pre- or postsynaptically in an activity-dependent, nonlinear fashion (Perea et al., 2009b). Accordingly, it would be interesting to characterize how this additional nonlinearity could ultimately shape synaptic learning rules (Brea et al., 2013; Porto-Pazos et al., 2011; Wade et al., 2011).

An interesting prediction comes from the consideration of the BCM rule for synaptic modification, according to which the threshold for LTP vs. LTD induction depends on postsynaptic activity (Bienenstock et al., 1982). Inasmuch as astrocytic gliotransmission could modulate such activity, then it could also change this threshold and thus be involved in metaplasticity (Hulme et al., 2014). In agreement with this prediction, in situations of morphological plasticity when the astrocytic coverage of synapses is reduced, the extracellular levels of D-serine get diluted, causing a decreased NMDAR activation which transforms stimuli expected to cause LTP into stimuli that elicit LTD instead (Panatier et al., 2006).

An important question in the induction of activity-dependent long-term plasticity is whether the temporal requirements of STDP can be satisfied in physiological conditions (Caporale and Dan, 2008; Sjöström et al., 2001). Indeed, timing and amplitude of EPSPs triggered by physiological spike trains are expected to dramatically regulate postsynaptic $Ca^{2+}$ entry (Froemke and Dan, 2002; Froemke et al., 2006, 2010), which is critical for the induction of long-term synaptic plasticity (Graupner and Brunel, 2010; Ismailov et al., 2004; Malenka et al., 1988; Mizuno et al., 2001a; Neveu and Zucker, 1996; Nevian and Sakmann, 2006; Yang et al., 1999a). On the other hand, because both these factors could be regulated by gliotransmission, acting either on the frequency (probability) of synaptic release or on the efficacy of postsynaptic receptors, it is plausible that astrocytes also play a role in the temporal requirements of STDP. The mechanism whereby this could occur is illustrated in Figure 4 by a phenomenological model for STDP (Graupner and Brunel, 2012; Higgins et al., 2014). The evolution of synaptic strength ensuing from postsynaptic $Ca^{2+}$ levels in response to a surrogate train of spikes (Figure 4A) is compared with the changes of synaptic strength triggered by the same spike pattern in the presence of short-term plasticity (Figure 4B) and, in addition, of modulation of synaptic release (Figure 4C) or NMDAR activation by gliotransmission (Figure 4D). In this model, the sign and entity of the variation of synaptic strength correlates with the average time spent by postsynaptic $Ca^{2+}$ above the thresholds for tLTD (*blue dashed line*) and tLTP induction (*orange dashed lines*) (Ismailov et al., 2004; Mizuno et al., 2001b; Nevian and Sakmann, 2006; Yang et al., 1999b). Modulation of $Ca^{2+}$ entry by short-term synaptic plasticity (Figure 4B) can alter this time window resulting in a different evolution of synaptic strength with respect to the case in which short-term plasticity is absent (Figure 4A) (Froemke and Dan, 2002; Froemke et al., 2006; Zucker and Regehr, 2002). Yet, the sign of the change of synaptic strength (i.e. tLTD or tLTP) could still be conserved depending on the pattern of presynaptic spikes (*left* vs. *right* panels). On the contrary, this may not be the case, in the presence of regulation of basal synaptic release or postsynaptic NMDAR activation by gliotransmitters. In these conditions, the sign of the variation of synaptic strength could be biased toward tLTD (or tLTP) or none (i.e. no plasticity), regardless of the presynaptic spike pattern (*asterisks* in Figures 4C,D). Moreover, additional stimulation could be required (*rightmost panels*) in order to observe any plasticity.

Although the role of astrocytes in setting the temporal requirements of STDP remains to be addressed, early experimental evidence supports the above theoretical prediction. At hippocampal synapses, induction of tLTP requires twice the number of pairings when endogenous glutamatergic gliotransmission is inhibited and synaptic release probability is decreased (Bonansco et al., 2011). This example suggests that gliotransmission could critically participate in STDP and possibly add further activity requirements for its induction. Intriguingly, in the hippocampus, where conflicting activity requirements for plasticity induction have been reported (Buchanan and Mellor, 2010), the existence of an astrocyte-mediated regulatory component, which was not previously considered, could possibly explain why the same induction protocol in different studies resulted either in no plasticity, or in tLTD or in tLTP (Buchanan and Mellor, 2007; Campanac and Debanne, 2008; Wittenberg and Wang, 2006), although other factors, such as differences in extracellular ion concentration could also likely contribute to the observed differences (Buchanan and Mellor, 2010).

### Astrocytes in heterosynaptic plasticity

In the majority of the studies hitherto reviewed, modulation of synaptic plasticity was induced either by exogenous stimulation of gliotransmission or in a homosynaptic fashion, by stimulation of the very synapses that were modulated by the astrocyte (Fossat et al., 2012; Henneberger et al., 2010; Jourdain et al., 2007; Kakegawa et al., 2011; López-Hidalgo et al., 2012; Min and Nevian, 2012; Navarrete and Araque, 2010; Panatier et al., 2006, 2011; Perea and Araque, 2007). Yet, astrocytes could also be implicated in the regulation of heterosynaptic plasticity, both of short- and long-term type.

In the hippocampus, endocannabinoids released by CA3-CA1 synapses trigger heterosynaptic potentiation via glutamate release from astrocytes (Navarrete and Araque, 2010). This heterosynaptic form of potentiation lasts only few minutes and is caused by an increase of neurotransmitter release probability in unstimulated synapses that are proximal to the activated astrocytes (Navarrete and Araque, 2010). On the other hand, at these same synapses, astrocytic glutamate has also been linked to an equally short-lived form of heterosynaptic depression, although in association with mild stimulation of Schaffer collaterals (Andersson et al., 2007a). Astrocyte-derived purines may also mediate heterosynaptic depression which, however, seems to last much longer: from few minutes (Zhang et al., 2003) to tens of minutes (Chen et al., 2013; Pascual et al., 2005; Serrano et al., 2006). Recent experiments (Chen et al., 2013) indeed showed that astrocyte-derived ATP is necessary and sufficient for classical heterosynaptic LTD that follows LTP induction at Schaffer collateral-CA1 hippocampal synapses (Abraham and Wickens, 1991; Lynch et al., 1977; Scanziani et al., 1996; Stanton and Sejnowski, 1989). The long-term nature of this heterosynaptic depression seems though dependent on the protocol of robust (tetanic) Schaffer collaterals stimulation deployed in those experiments (Chen et al., 2013), because mild stimulation was found instead to induce less durable (< 30 min) astrocyte-mediated heterosynaptic depression (Serrano et al., 2006; Zhang et al., 2003). It thus seems that both short- and long-lived forms of heterosynaptic plasticity mediated by astrocytes could coexist in the hippocampus, with their expression likely regulated by the characteristics of the inducing stimulus (Chen et al., 2013).

    Mediation of heterosynaptic plasticity by astrocytes may be regarded as a territorial form of regulation of plasticity, likely distributed across multiple synapses (Scanziani et al., 1996) within the astrocytes' anatomical domains, with potential, dramatic consequences in the learning and storage of information by the affected neural circuits (Stent, 1973; Willshaw and Dayan, 1990). In line with

the theoretical arguments previously exposed, heterosynaptic short-term plasticity mediated by astrocytes could transiently change the transmission mode across synaptic ensembles, favoring transmission of certain stimulus features with respect to others in an activity-dependent fashion, and accordingly, modulate STDP at these ensembles (Sjöström et al., 2008). In conditions of local deafferentiation, for example following focal ischemic injury (Murphy and Corbett, 2009), this mechanism could promote restoration of functional connectivity, endowing neural circuits with robustness to insults (Naeem et al., 2015; Wade et al., 2012). On the other hand, long-term modulation of synaptic strength within spatially defined astrocytic domains, could provide a substrate for synaptic clustering and competition (Larkum and Nevian, 2008; Song et al., 2005) which would be essential for nonlinear integration of synaptic inputs by dendrites (Legenstein and Maass, 2011) as well as network stability (Chistiakova and Volgushev, 2009).

**Astrocyte involvement in synaptic scaling and redistribution**

In the paraventricular nucleus of the hypothalamus, astrocytes are apparently responsible for an interesting form of heterosynaptic plasticity: they potentiate, likely via ATP release (Gordon et al., 2005), all the excitatory synapses of magnocellular neurosecretory cells, scaling them up in a multiplicative fashion (Gordon et al., 2009). This form of potentiation is peculiar because, like other forms of LTP, it can be observed within minutes from stimulation and last for tens of minutes, but it also shares some distinguishing features of classic multiplicative synaptic scaling (Turrigiano et al., 1998), as it increases all of a neuron's synapses by the same proportion in a non-Hebbian fashion. The scaling up indeed only requires presynaptic release of glutamate responsible for the astrocytic activation, but no postsynaptic activity (Gordon et al., 2009). Yet this type of scaling is different from conventional synaptic scaling presented in the Introduction, insofar as it potentiates synapses that are already active in a feedforward fashion, potentially increasing the probability that a stimulus successfully initiates firing. Remarkably, only synapses on cells that are proximal to the stimulated astrocytic processes are potentiated, suggesting that this enhancement of neural excitability is spatially confined to single neurons or small neuronal ensembles (Gordon et al., 2009). This could ultimately provide a mechanistic basis for the fast, finely-tuned release of hormones into the blood stream by magnocellular neurosecretory cells, which is a vital function of these cells (Brown et al., 2013).

Astrocytes however, may also participate in more traditional forms of homeostatic scaling (Pozo and Goda, 2010a; Turrigiano, 2006). In the hippocampus indeed, prolonged periods of activity deprivation result in synaptic upscaling that is mediated by an extracellular increase of TNFα derived from astrocytes (Beattie et al., 2002; Stellwagen and Malenka, 2006). Such TNFα increase regulates postsynaptic receptor trafficking, strengthening excitatory synapses while weakening inhibitory ones (Pribiag and Stellwagen, 2013; Stellwagen et al., 2005). In this fashion the overall network excitability is enhanced, thereby possibly compensating activity deprivation. As a homeostatic mechanism of plasticity, this astrocyte-mediated scaling rises within hours from activity deprivation and plateaus in a day or longer (Stellwagen and Malenka, 2006). However, differently from conventional scaling, it is not bidirectional (Turrigiano, 2006), meaning that, once established, it cannot be reversed by a decrease of extracellular TNFα (Stellwagen and Malenka, 2006). In addition, it does not interfere with the fast onset of activity-dependent LTP or LTD, raising the possibility of potential pathological instabilities of network activity in the long term (Savin et al., 2009; Volman et al., 2013). How to conciliate then this possibility with the fact that such instabilities were not observed experimentally (Beattie et al., 2002; Kim and Tsien, 2008; Stellwagen and Malenka, 2006)?

Although other mechanisms of regulation of synaptic plasticity that work in parallel with, and prevent instability by TNFα-mediated scaling are not to be excluded (Bains and Oliet, 2007; Pozo and Goda, 2010b), a possible answer to the above question comes from the consideration that TNFα is a special gliotransmitter (Santello and Volterra, 2012): at high concentrations it promotes massive glutamate release from astrocytes that may result in neurotoxicity (Bezzi and Volterra, 2001), but at constitutive extracellular concentrations, it is required for functional glutamate release from astrocytes (Santello et al., 2011). In this fashion, during activity deprivation, postsynaptic scaling by TNFα could coexists with glutamatergic gliotransmission and thus with the modulation of synaptic release probability by this latter and its possible effect on Hebbian learning. Consequently, synaptic strength could be the product of two contributions: (1) a synapse-specific Hebbian contribution, and (2) a synapse-nonspecific homeostatic contribution, both dependent on TNFα (Toyoizumi et al., 2014a). The ongoing homeostatic plasticity would allow scaling of maximal and minimal synaptic strengths set by Hebbian plasticity, constraining Hebbian modifications within a finite, activity-dependent range, and so making them inherently stable (Toyoizumi et al., 2014a). In this fashion, by means of TNFα release, and the regulation by this latter of activity-dependent glutamatergic gliotransmission, astrocytes could provide a unitary mechanistic substrate for the coordination of fast Hebbian learning and slow homeostatic plasticity that critically stabilizes neural circuit function (Turrigiano, 2008).

Is TNFα-mediated homeostatic plasticity the only mechanism whereby astrocytes could account for stabilization of Hebbian learning? Likely not: indeed, some of the previously reviewed mechanisms of plasticity that require astrocytes could be alternatively regarded as compensating mechanisms against Hebbian plasticity's inherent positive feedback. Heterosynaptic LTD mediated by hippocampal astrocytes observed in concomitance with LTP induction for example (Chen et al., 2013), could operate locally, by weakening synapses on the same dendritic branch of potentiated ones. This would allow keeping the overall activity in the dendrite constant while favoring synaptic competition and thereby stability (Rabinowitch and Segev, 2008; Van Rossum et al., 2000; Song et al., 2000). Long-term changes of synaptic release probability triggered by glutamatergic or purinergic gliotransmission (Deng et al., 2011; Halassa et al., 2009; Navarrete et al., 2012; Pascual, 2005; Schmitt et al., 2012), could also be regarded as mechanisms of synaptic redistribution (Markram and Tsodyks, 1996a). These forms of astrocyte-mediated plasticity would indeed ensue in long-term modulations of short-term plasticity that could condition the outcome of LTP or LTD upon the history of use of the synapse (Markram and Tsodyks, 1996b; Yasui et al., 2005), and possibly restrain the temporal window for STDP (Froemke and Dan, 2002; Sjöström et al., 2003; Volgushe et al., 1997). For example, strongly depressing synapses that most efficiently transmit transient increases in presynaptic firing, could instead turn better suited to transmit more sustained presynaptic firing by decrease of their short-term depression by purinergic gliotransmission (Halassa et al., 2009; Pascual, 2005; Schmitt et al., 2012; Zhang et al., 2003). On the other hand, in this scenario, larger firing rates may effectively be invoked to assure reliable transmission against the reduction in release probability that is concomitant with the decrease of short-term depression at these synapses (Abbott and Regehr, 2004). There are cases, however, where this increase of firing is undesirable, such as during wakefulness when the brain energy demand is high due to learning, but resources are limited (Buzsáki et al., 2002; Tononi and Cirelli, 2006). This scenario could be conciliated by the observation of a further mechanism whereby gliotransmission may modify synaptic plasticity. At synapses in the barrel cortex, the tonic activation of presynaptic A1 receptors (A1Rs) by astrocyte-derived adenosine, may directly promote postsynaptic NMDAR insertion through a sarcoma proto-oncogene tyrosine-protein (Src) kinase-dependent pathway (Deng et al., 2011). In this fashion, the decrease in synaptic release due to A1R activation would be counteracted, at least partly, by an increase of postsynaptic NMDA currents. Consequently, lower presynaptic rates would be needed to make the postsynaptic neuron fire despite the low reliability of the synapse. This would ultimately ensue in shorter latency of postsynaptic vs. presynaptic firing, favoring learning by tLTP while keeping low steady-firing rates (Abbott and Nelson, 2000), and so the energy demand (Fellin et al., 2012).

## Conclusions

Individual synapses are embedded in complex circuits in which many forms of plasticity operate. Multiple distinct mechanisms of plasticity may take place at the same synapse, or occur in different classes of synapses throughout the brain (Nelson and Turrigiano, 2008). While it seems likely that this diversity of mechanisms is tightly orchestrated so that each form of plasticity occurs in the right place at the right time to allow functionally appropriate tuning of neural circuits, such coordination of plasticity calls for the identification of possible candidate mechanisms for integration of neural activity and initiation of the plastic changes. The evidence reviewed here expands our knowledge of plasticity, adding astrocytic gliotransmission as a further possible determinant for it, and points to astrocytes as potential integrators in space and time of neural activity and its plasticity. Astrocytes are indeed well positioned to sense afferent inputs and ideally suited to temporally and spatially integrate synaptic signals in order to change the efficacy of many synapses. On the other hand the dendritic arborization of a single neuron could be covered by tens of different astrocytes (Halassa and Haydon, 2010). Hence, that single neuron and its dendritic arbor may be influenced by several different astrocytes. This expands our notions of synaptic and neural domains since the synapses of a neuron would not only convey signals to other neurons but also interact with multiple astrocytes, and, vice versa, a single astrocyte could interact with synapses of multiple neurons. In this fashion, astrocytes could bridge and influence neural circuits that are not directly connected, participating in the formation of complex neural ensembles that respond on a large variety of temporal and spatial scales. At the same time, being able to sense neural activity impinging on them from different regions of the neural circuitry, astrocytes could ideally orchestrate plastic changes at different spatial and temporal scales, in order to ensure the stability and function of circuits.

# MODELLING BOX: CONDITIONS FOR ASTROCYTE REGULATION OF SHORT-TERM SYNAPTIC PLASTICITY

According to the classical quantal hypothesis of synaptic transmission, the average postsynaptic response PSC may be thought as the product of the postsynaptic quantal size $q$, the number $n$ of release sites and the probability $p$ that release of a quantum of neurotransmitter occurs at a site, i.e. PSC = $n\,p\,q$ (Del Castillo and Katz, 1954). The probability $p$, on the other hand, may be computed as the product of the probability $x$ that a neurotransmitter-containing synaptic vesicle that is available for release by the probability $u$ that such vesicle could be docked and thus be effectively released, i.e. $p = u\,x$ (Südhof, 2004; Zucker and Regehr, 2002). Following (Tsodyks et al., 1998), the average temporal evolution of $u$ and $x$ as a function of the instantaneous rate $v(t)$ of incoming action potentials may be described by the simple set of ordinary differential equations

$$\frac{du}{dt} = \frac{b-u}{\tau_f} + b(1-u)\cdot v(t)$$
$$\frac{dx}{dt} = \frac{1-x}{\tau_d} - u\cdot x\cdot v(t)$$

where $b$ represents the synaptic release probability at rest and is related to the resting intrasynaptic $Ca^{2+}$ concentration while $\tau_d$ and $\tau_f$ respectively denote the rate of synaptic depression and facilitation. By simple algebraic calculations, it is possible to solve the above equations for the steady state (i.e. $du/dt = dx/dt = 0$) and see that there exists a critical value $b_\theta = \tau_f / (\tau_d + \tau_f)$ such that only if $b < b_\theta$ the synapse may exhibit facilitation depending on the rate $v$, otherwise synaptic transmission is dominated by short-term depression (Tsodyks, 2005).

In the presence of gliotransmission from the astrocyte, intrasynaptic $Ca^{2+}$ levels and thus the synaptic release probability $b$, may be thought to depend on the fraction $\gamma$ of extrasynaptically-located presynaptic receptors bound by the gliotransmitter, i.e. $b = b(\gamma)$. In absence of quantitative physiological data, the function $b(\gamma)$ may be assumed to be analytic around zero and its first-order expansion be considered accordingly, i.e.

$$b(\gamma) \simeq b(0) + \left.\frac{db}{d\gamma}\right|_{\gamma=0} \gamma + O(\gamma^2)$$

The zeroth order terms $b(0) = b_0 = const$ must correspond to the value of $b$ in the absence of gliotransmission. The first order term instead may be assumed to be linear in $\gamma$, at first instance for the sake of analytical tractability, such as $\left.\frac{db}{d\gamma}\right|_{\gamma=0} = -b_0 + \alpha$, where the parameter $\alpha$ varies between 0 and 1 and accounts for the nature of bound presynaptic receptors: release-decreasing for $\alpha < b_0$, and release-increasing for $\alpha > b_0$ (De Pittà et al., 2011). Assuming that each receptor may exist in only two states (bound/not bound with probability $\gamma$ and $1-\gamma$ respectively), on average, the fraction of bound receptors for an instantaneous rate of gliotransmitter release from the astrocyte $\psi(t)$ may be written as (De Pittà et al., 2011):

$$\frac{d\gamma}{dt} = -\frac{\gamma}{\tau_g} + Jg(1-\gamma) \cdot \psi(t)$$

where $g$ denotes the probability of gliotransmitter release from the astrocyte and $J$ represents the fraction of released gliotransmitter that effectively binds to presynaptic receptors, while $\tau_g$ is the decay time for the modulatory effect on synaptic release mediated by the receptor. In the simplest scenario, $g$ may be thought to be proportional by a constant factor $a$ to the fraction of astrocytic gliotransmitter resources available for release for which an analogous equation to that of $x$ may be written, once $\tau_d$ is replaced by $\tau_a$, i.e. the rate of reintegration of released gliotransmitter, and $\nu(t)$ by $\psi(t)$.

The threshold rate $\psi_\theta$ for gliotransmitter release for which a change of paired-pulse plasticity could be observed in experiments is then defined as the steady-state rate of gliotransmitter release from the astrocyte for which the synaptic release probability at rest equals the threshold value for the switch between facilitation and depression, i.e. $b(\psi_\theta) = b_\theta$, and reads:

$$\psi_\theta = \frac{b_\theta - b}{a\left[(b-b_\theta)\tau_a + J(\alpha - b_\theta)\tau_g\right]}$$

Because it must be $\psi_\theta \geq 0$, mapping of $\psi_\theta$ on the plane $b$ vs. $b_\theta$ for different values of the three parameters $J$, $\tau_a$, $\tau_g$ and all possible allowed values of $\alpha$ ultimately provides the maps in Figure 2.

The above expression for $\psi_\theta$ holds true only if the dynamics of gliotransmitter-mediated modulation (as reflected by the decay time $\tau_g$) is sufficiently slower than gliotransmitter reintegration (on the time scale $\tau_a$) and synaptic dynamics (characterized by the time constants $\tau_d$ and $\tau_f$), so that the latter can be assumed to be at steady state (De Pittà et al., 2011). This seems indeed the case since typical synaptic time constants are estimated in the range of hundreds of milliseconds to few seconds (Markram et al., 1998b), and re-acidification, and possibly re-filling, of gliotransmitter-containing vesicles could be as fast as $\tau_a \approx 1.5$ s (Bowser and Khakh, 2007; Marchaland et al., 2008). On the other hand, the modulation of synaptic release by gliotransmitters could last from tens of seconds to minutes whether it is mediated by glutamate (Andersson et al., 2007b; Araque et al., 1998a, 1998b; Fiacco and McCarthy, 2004; Jourdain et al., 2007; Liu et al., 2004b; Navarrete and Araque, 2010; Perea and Araque, 2007; Perea et al., 2014; Perez-Alvarez et al., 2014) or purine receptors (Pascual et al., 2005; Serrano et al., 2006; Zhang et al., 2003).

The fraction $J$ of gliotransmitter molecules bound by presynaptic receptors characterizes the coupling of the astrocyte with the synapse and may be thought to depend on a multitude of parameters such as the source of gliotransmitter, the spatiotemporal features of gliotransmitter release and clearance in the extracellular space, the affinity of presynaptic receptors for gliotransmitter molecules, and the subcellular localization of these receptors in relation to the source of gliotransmitter (Araque et al., 2014; Hamilton and Attwell, 2010). Because only few of these parameters can be accurately measured, and most will likely vary from synapse to synapse, any attempt to estimate $J$ must therefore focus on an average behavior while bearing in mind the possible sources of variability of the latter. Remarkably, the time course for the clearance of purines in the extracellular space may be considerably slower than that of glutamate, due to the possible conversion of one purine into another one by ectonucleotidases like, for example, the degradation of ATP into adenosine, which could bind additional presynaptic receptors (Abbracchio et al., 2009). Thus, for equal coupling conditions, $J$ could be larger for purinergic than glutamatergic gliotransmission, which could also explain why purines, rather than glutamate from astrocytes, could induce tonic activation of presynaptic receptors (Fellin et al., 2009; Halassa et al., 2009; Schmitt et al., 2012). Indeed, a slow clearance could facilitate the extracellular accumulation of purines at multiple synaptic terminals surrounding the release site in the astrocyte, thereby setting the tone of activation of presynaptic receptors. In the simulations in Figure 2B, Figure 3 and Figure 4 we set $J = 0.8$ along with $g = 0.6$, $\tau_a = 1.25$ s, $\tau_g = 5$ s, $\tau_d = 0.5$ s, $\tau_f = 1$ s and $b_0 = 0.1$ (Figure 3A) and $b_0 = 0.9$ (Figure 4B).

**Acknowledgments**

MDP wishes to thank S. H. R. Oliet, A. Araque and J. D. Bains for illuminating conversations. MPD acknowledges support by a Maria Skłodowska-Curie International Outgoing Fellowship (Project 331486 "Neuron-Astro-Nets"). Work in AV lab is supported by the ERC Advanced grant 340368 "Astromnesis" and by grants from the Swiss National Science Foundation (SNSF) 31003A–140999, NCCR "Synapsy" and NCCR "Transcure".

**Figure 1**. **Gliotransmitter-mediated pathways for regulation of synaptic transmission**. Synaptically released neurotransmitters (1) that spill out of the synaptic cleft (and are not sequestered by transporters), and retrograde messengers like endocannabinoids (2), can bind astrocytic GPCRs and transiently elevate cytosolic $Ca^{2+}$ in the astrocyte (3). In a similar fashion, neuromodulators such as acetylcholine or noradrenaline (4) may also trigger astrocytic GPCR-dependent $Ca^{2+}$ signaling. Other pathways for activity-dependent increases of astrocytic $Ca^{2+}$ that do not involve GPCRs could also exist (Verkhratsky et al., 2012). An increase of cytosolic $Ca^{2+}$ may promote release of gliotransmitters from the astrocyte (5) which could affect synaptic transmission both pre- and postsynaptically. Presynaptically, gliotransmitters like glutamate, ATP or its derivate adenosine, bind extrasynaptic receptors that modulate the release probability of synaptic neurotransmitters (6). Postsynaptically, astrocyte-derived D-serine controls the degree of activation of NMDARs at cortical and hippocampal glutamatergic synapses (7). On the other hand, in several brain regions, ATP, glutamate, GABA or TNFα released from astrocytes could also bind postsynaptic (8) and extrasynaptic receptors (9), directly contributing to postsynaptic depolarization and/or receptor trafficking. Remarkably, ambient TNFα, perhaps of astrocytic origin, seems a permissive factor in the release of other gliotransmitters like glutamate (10), suggesting that both pre- and postsynaptic signaling pathways mediated by gliotransmitters may require specific conditions to occur.

**Figure 2**. **Conditions for modulation of short-term plasticity by gliotransmitters**. **A** For given synaptic time constants, a synapse could show either short-term facilitation (*green region*) or short-term depression only (*orange region*) depending on whether its probability of neurotransmitter release at rest ($b$) is below or above a threshold value $b_\theta$ (*white line*). In terms of paired-pulse plasticity, this threshold corresponds to a paired-pulse ratio (PPR) equal to one, while short-term depression and facilitation are characterized by PPR values respectively smaller and larger than one (Zucker and Regehr, 2002). In principle, all facilitating (respectively, depressing) synapses could turn depressing (respectively,

facilitating) by an appropriate increase (respectively, decrease) of *b* across this threshold, mediated by presynaptic receptors (*yellow arrows*). **B** In practice, when such receptors are activated by gliotransmitters, only a subset of synapses (*colored regions* between *magenta dotted lines*) can effectively show a change in short-term plasticity, depending on the rate of gliotransmitter release from the astrocyte. The size of the parameter region for which a qualitative change can happen dramatically reduces when: **C** the reintegration of released gliotransmitter by the astrocyte is too slow, or **D** the modulation of synaptic release is weak, or **E** the morphology of the astrocyte-synapse coupling does not allow efficient activation of presynaptic receptors (e.g., when the astrocytic process is not close enough to the presynaptic terminal of the synapse), or for a combination of any of these three factors. Therefore, a modulation of short-term plasticity by the astrocyte may not be detectable despite the existence of gliotransmission. Plots were generated by a model of astrocyte-regulated synapses described in the Modelling Box. Figures in **C**-**E** use the same model parameters of **B** except for **C** a two-fold increase of $\tau_a$; **D** a 2.5-fold reduction of $\tau_g$, and **E** a four-fold decrease of *J*.

**Figure 3**. **Modulation of synaptic release probability by gliotransmitters may correlate with a change in the frequency response of the synapse**. **A** Synapses with intermediate values of release probability are typically characterized by a bell-shaped frequency response (*black curve*) whereby they most effectively transmit incoming presynaptic action potentials within an intermediate frequency band. **B** This behavior could however be altered by an increase of synaptic release probability upon activation of presynaptic receptors by gliotransmitters. In this scenario for example, the synapse's frequency response could become monotonically decreasing (*red curve*), akin to a low-pass filter. Each point in **A**, **B** represents the simulated fraction of neurotransmitter released from the synapse by a 100 s-long Poisson train of actions potentials at rate *v*, averaged across 500 trains. **C** A change in the synapse's frequency response mediated by gliotransmitters is expected to dramatically alter how presynaptic firing rates (*top panel*) are transmitted (*lower panels*, Postsynaptic Currents).

Simulated postsynaptic currents were obtained from the fraction of synaptic neurotransmitter that was released by Poisson trains of action potentials with a sequence of instantaneous transitions from 0 to 32 Hz (*top panel*). To generate realistic postsynaptic AMPAR currents, the fraction of synaptically released neurotransmitter was convolved by an α-function with time constant 10 ms (Van Vreeswijk et al., 1994). Each PSC trace was averaged across 500 trials. Gliotransmitter release was assumed to occur regularly every 2 s and activate at least 90% of presynaptic receptors (see Modelling Box for further details).

**Figure 4**. **Potential mechanisms of modulation of STDP by gliotransmitters**. **A** In a $Ca^{2+}$-based description of STDP (Graupner and Brunel, 2012; Higgins et al., 2014), the timing of pre- vs. postsynaptic action potentials (*top panels*) modulates postsynaptic intracellular $Ca^{2+}$ (*middle panel*) which in turn drives long-term modification of synaptic strength (*bottom panel*). An increase in synaptic strength occurs whenever postsynaptic $Ca^{2+}$ is above the threshold for LTP (*orange dashed line* and *shades*), whereas a decrease happens when $Ca^{2+}$ is below this threshold but above the one for LTD (*blue dashed line* and *shades*). In this fashion the model allows single spikes to change synaptic strength but the magnitude of this change becomes significant only for multiple opportunely timed spikes in agreement with classic protocols of STDP induction (Graupner and Brunel, 2012). **B** Short-term depression decreases synaptic release with respect to its basal value (*red dashed line*) in an activity-dependent fashion, and in turn decreases postsynaptic $Ca^{2+}$ entry due to presynaptic action potentials. This may alter the variation of synaptic strength with respect to the case in **A** without short-term plasticity, so that the same pre/post action potentials result, for example, in weaker LTP. **C** Postsynaptic $Ca^{2+}$ entry could also be regulated by modulations of synaptic release probability via gliotransmitters. Thus, for a reduction of synaptic release probability (*red arrow* to *green dashed line*) mediated by gliotransmitter-bound presynaptic receptors, stimuli that would induce LTP could result in LTD instead (*red asterisks*). **D** An increase of postsynaptic $Ca^{2+}$ entry due to a larger number of NMDARs activated by presynaptic spikes in the presence of increased extracellular concentration of astrocyte-derived D-serine could reverse the outcome of plasticity, but this would also depend on the exact induction stimulus. In these conditions indeed, only some portions of the stimulus could induce, or even enhance LTP, as in the case without gliotransmission (*green asterisk*), but this potentiation could be canceled by subsequent pre/post spikes. Consequently, no net plastic change could be observed (*black asterisk*), and further stimulation would be required to induce any potentiation (rightmost panels for $t > 3$ s). **B** Short-term depression was implemented as described in the Modeling Box with a basal synaptic release probability of 0.9. **C** The modulation of synaptic release by gliotransmitters was mimicked by a 60% reduction of this release probability. **D** The regulation of NMDAR efficacy by astrocytic-derived D-serine was mimicked instead by a 65% average increase of postsynaptic $Ca^{2+}$ entry per presynaptic action potential.

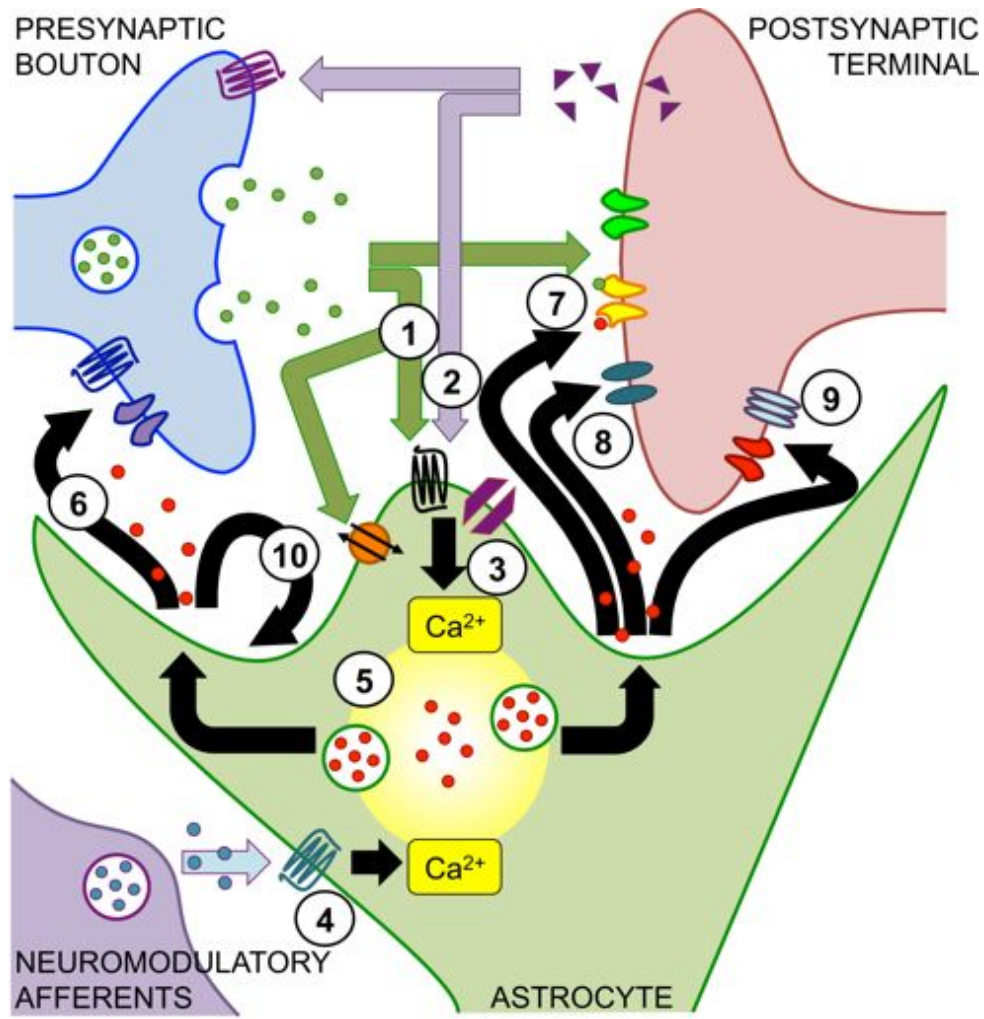

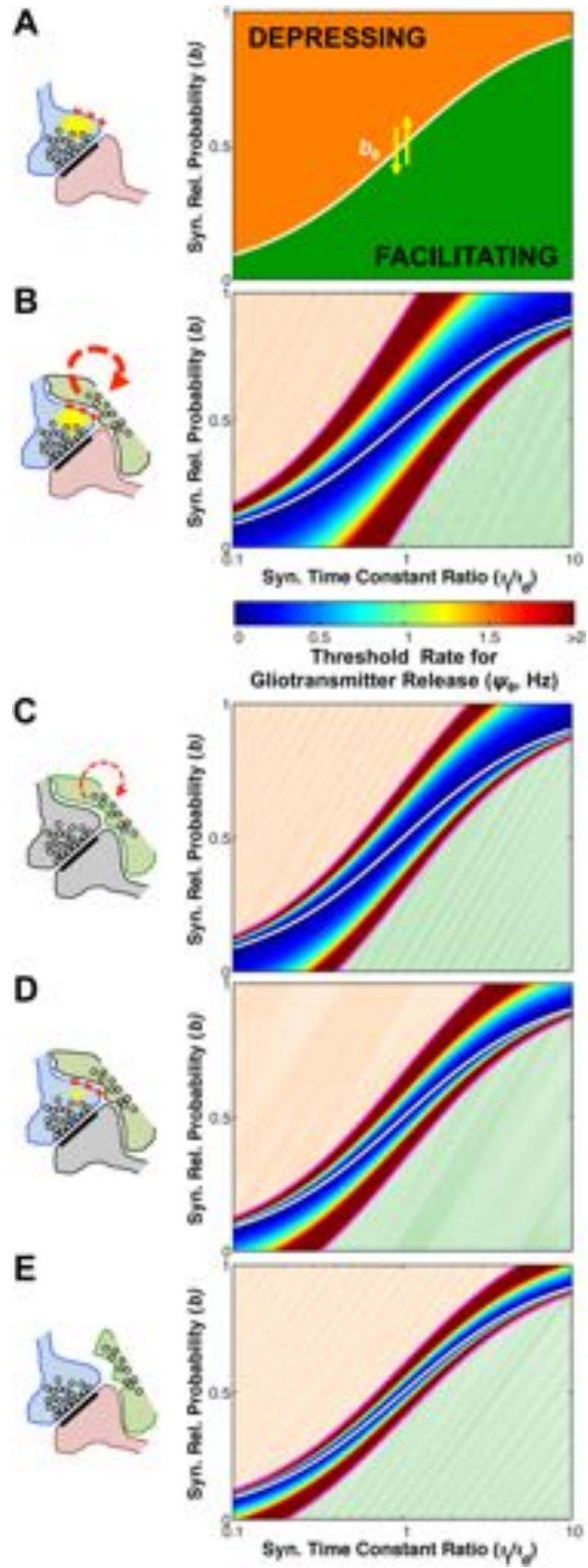

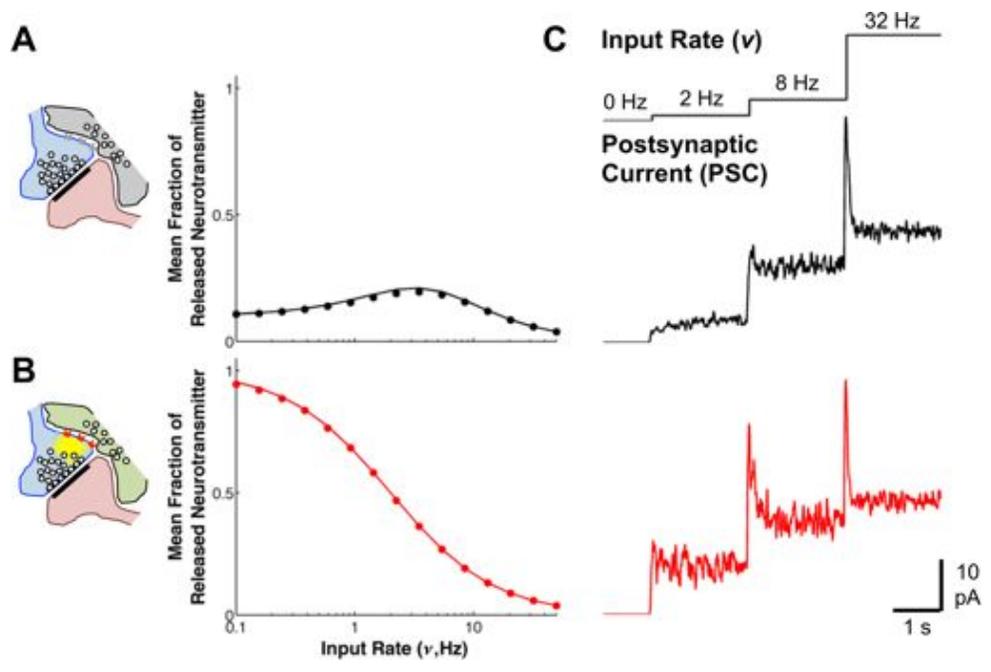

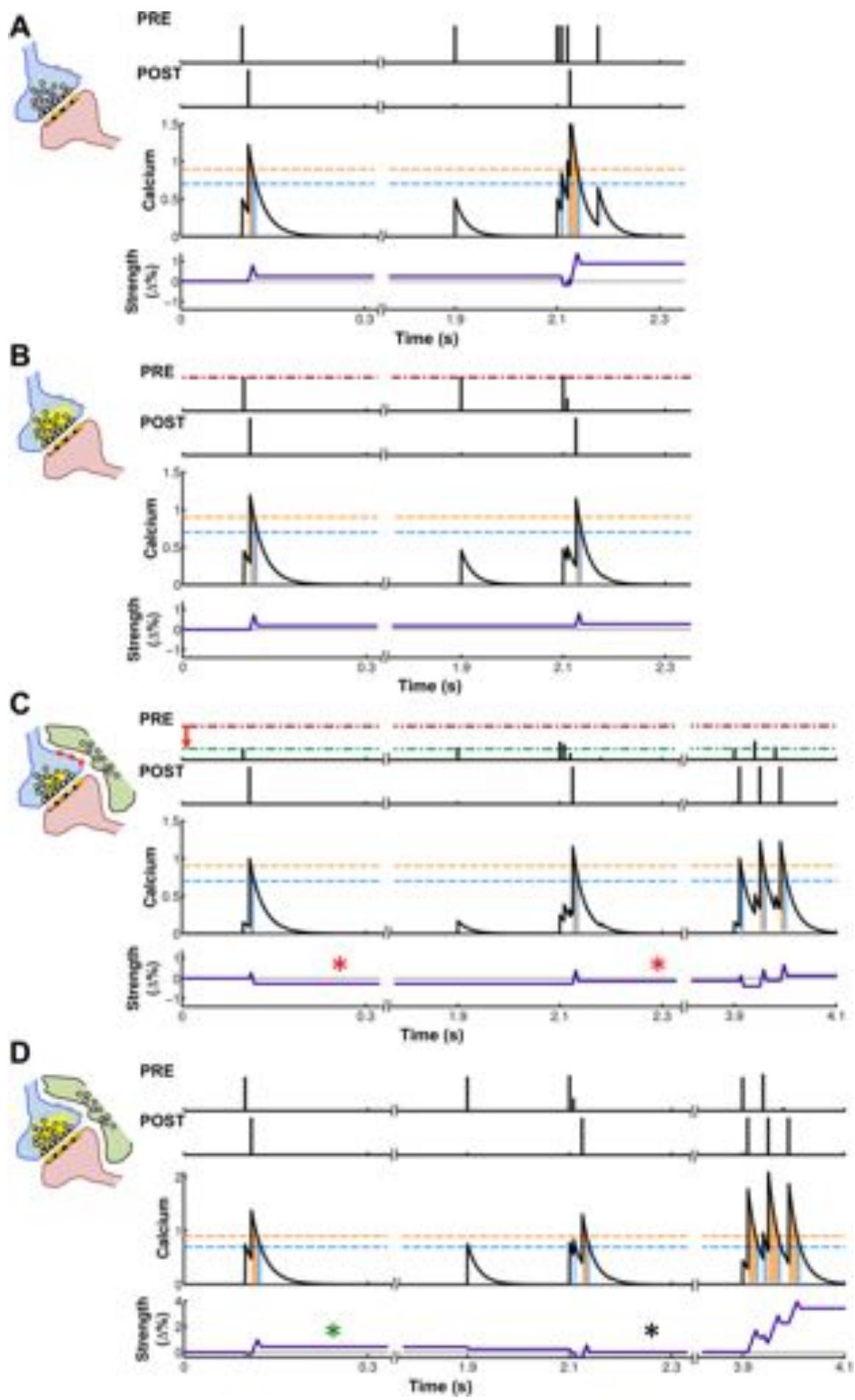